\documentclass{article}

\usepackage{arxiv}

\usepackage[utf8]{inputenc} 
\usepackage[T1]{fontenc}    
\usepackage{hyperref}       
\usepackage{url}            
\usepackage{booktabs}       
\usepackage{amsfonts}       
\usepackage{nicefrac}       
\usepackage{microtype}      
\usepackage{bbold}
\usepackage{amsmath}
\usepackage{mathtools}
\usepackage{dsfont}
\usepackage{tabu}
\usepackage{algorithmic}
\usepackage{algorithm2e}
\usepackage{setspace}
\usepackage[all]{background}
\backgroundsetup{contents={}}

\title{Privacy-Aware Distributed Mobility Choice Modelling over Blockchain}

\author{David L\'opez\\
  Grupo de Investigación en Ingeniería de Transporte y Log\'istica \\
  Universidad Nacional Aut\'onoma de M\'exico\\
  Mexico City, Mexico \\
  \texttt{dlopezfl@iingen.unam.mx} \\
   \And
 Bilal Farooq \\
  Laboratory of Innovations in Transportation\\
  Ryerson University\\
  Toronto, Canda \\
  \texttt{bilal.farooq@ryerson.ca} \\
}

\begin{document}
\maketitle

\begin{abstract}
A generalized distributed tool for mobility choice modelling is presented, where participants do not share personal raw data, while all computations are done locally. Participants use Blockchain based Smart Mobility Data-market (BSMD), where all transactions are secure and private. Nodes in blockchain can transact information with other participants as long as both parties agree to the transaction rules issued by the owner of the data. A case study is presented where a mode choice model is distributed and estimated over BSMD. As an example, the parameter estimation problem is solved on a distributed version of simulated annealing. It is demonstrated that the estimated model parameters are consistent and reproducible.
\end{abstract}

\section{Introduction}
With the advances in ubiquitous networks, nowadays it is possible to obtain detailed information on individual person's behaviour by using smartphones, cellphone towers, Wi-Fi hot-spots or traffic sensors~\cite{farooq2015ubiquitous}. All this information could potentially boost the accuracy of prediction tools---especially in the field of mobility choices modelling. Nowadays, data sources are vast and literally every individual with a smartphone is a potential data source.

Protecting the privacy of individuals is becoming more and more important. Privacy-preserving is a major challenge when sharing information to third parties. While some researchers or agencies may use private information in a responsible way others could misuse the private information for shady purposes~\cite{Cadwalladr2018}. On the other hand analyzing private information from a large scale population can bring great benefits to humankind as it can help providing better understanding on health care~\cite{Raghupathi2014}, mobility~\cite{Badu-Marfo2019a}, and other fields~\cite{Amalina2019}.

With the use of distributed ledger technologies like blockchain, it is possible to create secure networks. In broad terms, blockchain is distributed ledger where participants can write or query their contents in a transparent manner. The first iteration of blockchain was specifically designed to make the participant anonymous~\cite{Nakamoto2008}. However, when dealing with people sharing information it is desirable that not all participants of the blockchain are anonymous. In order to build trust with people, they may need to know with whom the information is been shared. Furthermore, there has to be some level of control over the nodes in terms of who can participate in consensus mechanics, so they cannot accumulate enough computational power to tamper the blockchain~\cite{Bach2018}. With these issues in mind a multi-layered blockchain framework over the \emph{public closed} Blockchain for Smart Mobility Data-market (BSMD) was developed by~\cite{Lopez2018}, where participants own and shares their mobility data while their privacy is preserved. The extensive details of the framework can be found in~\cite{lopez2019multi}.

Choice (e.g. mode, route, start-time) modelling is an important aspect of mobility demand analysis where detailed disaggregate data is collected from individuals and used to develop parametric models inspired from microeconomics and econometric. The individuals have no control on the data they provided and anyone having access to the collected data can use their information without their knowledge. Using the basic principles of distributed ledger technologies and BSMD framework, we propose an alternative for mobility choice modelling, which is able to estimate parameters on distributed data from different sources where:
\begin{enumerate}
    \item Personal raw information (e.g. age, gender, income, etc.) are never shared
    \item Computations are always done locally
    \item No central cloud servers are required to store the information
\end{enumerate}
Users are always in control of their information as no raw personal information is shared and all the information transactions are recorded on the blockchain. Thus the users can track where their information is used.

In this paper a choice modelling tool distributed over the BSMD is proposed. The main objective is to develop an environment where participant can run choice models using personal information, but where personal raw data is never shared. The rest of the paper is organized as follows: it is first introduced the background on how blockchain has been used for distributing process and databases while maintaining the privacy of individuals. We develop a methodology for creating distributed models and share results over the blockchain. A case study is presented on a stated preference travel survey with mode choice information. Privacy and security aspects are discussed. In the end, concluding remarks and future work are presented.

\section{Related work}
Mobility choice modelling is been an active field in the past decades and it has evolved from standard logit or probit models~\cite{Barff1982} to more detailed models that can account for heterogeneity and dynamics~\cite{Hagenauer2017}. To mention a few recent examples, mode choice was studied from the perspective cultural context~\cite{Ramezani2018}, social influence~\cite{Pike2018} and crime perception~\cite{Halat2015}.

The common denominator of choice modelling is that the data that feeds the models is centralized which raises privacy concerns. Also, accessing big amounts of data from heterogeneous groups is always desirable for creating richer models, but this is often expensive and out of reach for entities like non-for-profit organizations. As of today, to the best of our knowledge the privacy and cost problems have not been tackled specifically in the field of choice modelling.

In recent years, blockchain networks have gained attention because of their ability to transact assets securely and privately. This has resulted in blockchain applications that are outside the world of cryptocurrencies. In recent developments, blockchain has been used for distributed databases~\cite{Muzammal2019} and crowdsourcing~\cite{Li2019}, where participants can query or write the ledger. But also blockchain have been used to distribute process while preserving the privacy and security of individuals, distributed applications over the blockchain have been created for IoT applications~\cite{Stanciu2017}, smart grids management~\cite{Pop2018} and artificial intelligence ~\cite{Yang2019}.

In this paper we explore how choice modelling can be distributed over a blockchain, while privacy is maintained and raw personal data do not necessarily have to be shared. In particular, we tested our proposed methodology for the \emph{maximum likelihood} based parameter estimation of choice model using a distributed version of simulated annealing on BSMD framework. The objective of the paper is to show how to distribute a process over the BSMD where the personal information used to feed the models is always in control of the individual who generate such information.
\section{Distributed behavioural choice modelling over the BSMD}
\label{sec:distributed_bsmd}
This study developed a distributed behavioural choice modelling framework over BSMD proposed in~\cite{Lopez2018, lopez2019multi}. The BSMD is a \emph{public blockchain} meaning that the blockchain ledgers is open for consultation but a special permission is need it participation in consensus mechanisms~\cite{Olnes2017}. 

Nodes in the BSMD owns a digital identification (\emph{id}) where participant locally store a wide range of personal information such as sensing mobile data, state preferences or even mathematical models. Node may opt to add in \textit{id}  and \textit{identity key} which issued by trusted agency (e.g. government or non-profit organization). This key is used by other nodes to verify claims, for instance, if a node claims that they are over 18years old, other nodes can use their identity key and asks the issuer if the node is telling the truth.

To self-enforce fair trade in transactions of information \emph{smart contracts} are activated before all transactions. In these the participants set the terms for sharing information. In a distributed process the participants need to set at least the following terms: (1) temporality (how long the participant are willing to participate in the distributed process, (2) idle (nodes may opt to participate only when his device is idle) and (3) permissions (nodes could select to publicly share information in the blockchain).

When a node wants to start a distribute process for a behavioural choice model, it must first create a \emph{domain} and announce it over the blockchain so that participant nodes can join. In the announcement the node preset itself as the \emph{chief} and clarifies the type of observations it needs to estimate the choice model and the incentives for participating in the model estimation process. Once a node joins the \emph{domain} it is registered as a \emph{worker} node and the \emph{chief} node can solicit its services for the period established in the terms of the \emph{smart contract}.

\begin{figure*}[htbp]
    \centering
    \includegraphics[width=.90\textwidth]{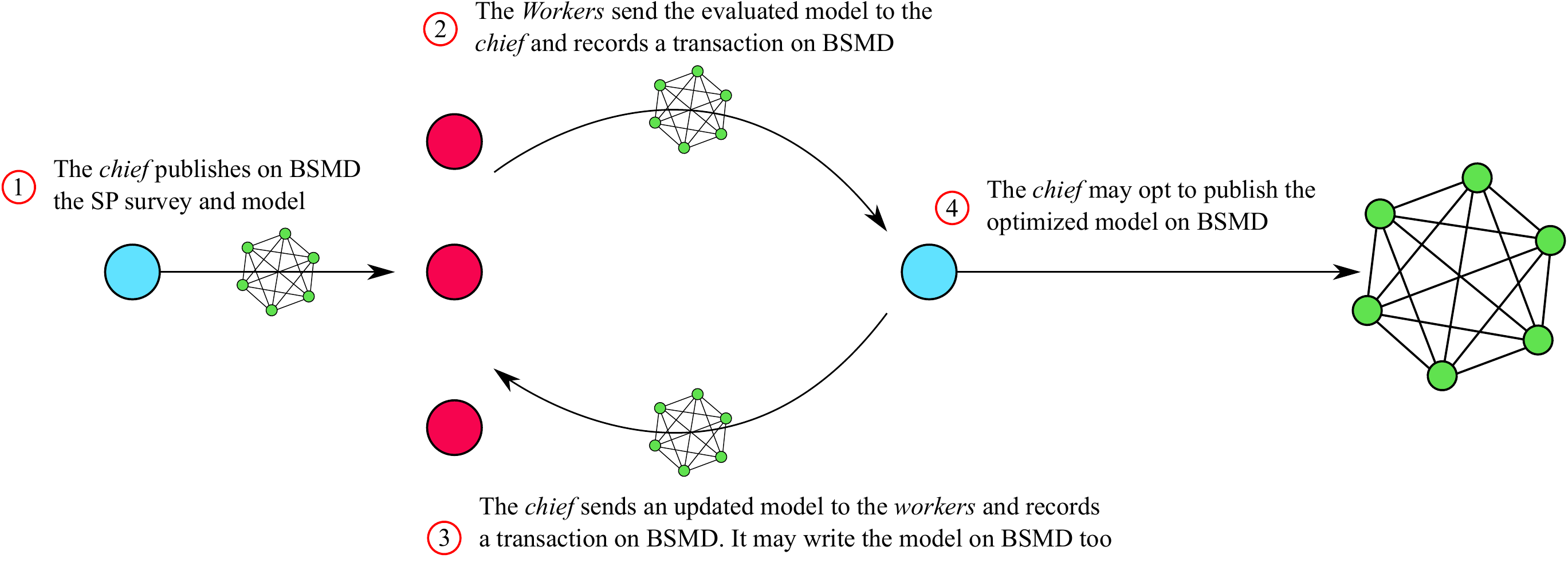}
    \caption{Distributed behavioural choice modelling over the BSMD}
    \label{fig:chioce_model_BSMD}
\end{figure*}

The distributed process for choice modelling is depicted in Fig.~\ref{fig:chioce_model_BSMD} and it starts when the \textit{chief} sends a connection request to each worker. The request includes terms defined by the \textit{chief} and \textit{worker} that will be used in the \textit{smart contract}. If the terms of all involved parties do not contradict each other, a \textit{peer-to-peer} channel of communication is opened between the chief and each worker for sharing information. The \emph{chief} node then publishes in the ledger a state preference (SP) survey and an objective function, e.g. \textit{log-likelihood} function in case of \textit{max-likelihood} method (see step 1 in Fig.~\ref{fig:chioce_model_BSMD}). The \emph{workers} answer the survey and keep their responses on their devices. In the following step, \textit{worker} nodes uses their responses to evaluate the objective function and send back their results to the \emph{chief} (see step 2 in Fig.~\ref{fig:chioce_model_BSMD}). 

It is important to note that \emph{workers} nodes send the evaluated function to the \emph{chief} using a \emph{peer-to-peer} connection where the information is encrypted. Although the transaction details such as sender, receiver, date, etc are written in the ledger the evaluation value of the objective function is not--publishing such values in the ledger may result in the possibility to reverse-engineer the evaluation value and infer the SP survey response on an individual.

Once the \emph{chief} collects and merge all the objective function values, if necessary, it may send the new model to the \emph{workers} for another round and the procedure starts again (see step 3 in Fig.~\ref{fig:chioce_model_BSMD}) until the results are optimized. The \emph{chief} sends models using the \emph{peer-to-peer} connection and it may write the model in the ledger if it wants to do so (see step 4 in Fig.~\ref{fig:chioce_model_BSMD}). Since objective functions from single participants are merged it is harder to extract personal information from the model. 

The use of \emph{peer-to-peer} connections for data transactions in the BSMD ensures the protection of the privacy of individuals, since some information is privacy-sensitive and it cannot be published on the ledger. Furthermore, the amount of information that can be contained in a single transactions are limited by the size of the blocks forming the blockchain. Hence, if the models or evaluations are too complex and are of large size, it is better to directly send the information. If nodes want to publish complex models they can use cloud serves and publish in the ledger the address of their optimized models. In the next section, we demonstrate a distributed version of simulated annealing (a parameter estimation method) over the BSMD to evaluate behavioural choice modelling.

\section{Case Study}
The distributed behavioural choice modelling over BSMD is demonstrated using the maximum \emph{likelihood} method to estimate the parameters for a mode choice model over a stated preference (SP) choice survey by~\cite{Sobhani2017}. For model evaluation a distributed version of simulated annealing~\cite{ross1990course} is used, however any other parameter estimation algorithm can be implemented with minimum effort. The SP survey was collected in 2017 consisting of commuters along the rail corridor between Montr\'eal in Canada and NYC, Maine, and Boston in USA. In the SP choice survey, each respondent was presented with up to 6 alternatives: automobile, automobile rental, bus, plane, train, and train-hotel, where each mode alternative is characterized with trip duration, trip reliability and trip cost. For testing the distributed choice model we use a subset of the SP survey consisting in $N = 246$ observations, only including automobile and train as the two mode choices.

The choice model of user preferences between train ($tr$) and automobile ($a$) is computed with the \emph{likelihood} function associated with the binary \textit{logit} model as shown in Equation~\ref{eq:likehoodmodel}. The parameters $\beta_{a}$, $\beta_{c}$ and $\beta_{t}$ are associated with average automobile-preference, cost and travel time respectively. In Equation~\ref{eq:v_cost} the variables $c_{a,n}$ and $c_{tr,n}$ are associated to automobile and train costs of the observation $n$ while $t_{a,n}$ and $t_{tr,n}$ are associated to automobile and train travel times of the observation $n$.

\begin{equation}
\label{eq:likehoodmodel}
\max_{\vec{\beta}} \sum_{n=1}^{N} \ln((y_{a,n}P(a_n;\vec{\beta}) + y_{tr,n}P(tr_n;\vec{\beta})))
\end{equation}
\noindent
where,
\begin{align*}
y_{a,n}     & = 1 \text{ if the observation $n$ is automobile, else } 0 \\
y_{tr,n}     & = 1 \text{ if the observation $n$ is train, else } 0 \\
\vec{\beta} & = (\beta_{a}, \beta_{c}, \beta_{t}) 
\end{align*}
\begin{align}
\label{eq:logitmodel}
P(a_n;\vec{\beta})   &= \frac{e^{V_{\vec{\beta},a_n}}}{1 + e^{V_{\vec{\beta},a_n}}} \\
P(tr_n;\vec{\beta})  &= 1 - P(a_n;\vec{\beta}) \\
\label{eq:v_cost}
V_{\vec{\beta},a_n}  &= \beta_{a} + \beta_{c}(c_{a,n} - c_{tr,n}) + \beta_{t}(t_{a,n} - t_{tr,n})
\end{align}

The behavioural choice model of Equation~\ref{eq:likehoodmodel} is obtained by using a distributed simulated annealing (dSA) algorithm over one \emph{chief} and four \emph{worker} nodes. Algorithm~\ref{alg:chief} and~\ref{alg:worker} show the pseudo code of the dSA hosted in the \emph{chief} and \emph{workers} nodes respectively. The dSA algorithm starts when the \emph{chief} distributes the $\vec{\beta}$ parameters among the \emph{worker} nodes (see lines 2--4 in Algorithm~\ref{alg:chief}). Then each \emph{worker} node use his personal observations and $\vec{\beta}$ to compute his personal choice, $l_w$, and send $l_w$ to the \emph{chief} node (see Algorithm~\ref{alg:worker}). Next, the \emph{chief} collects and summarize all $l_w$'s from each worker to compute a global choice $l$ (see line 5 in Algorithm~\ref{alg:chief}).

Once the initial parameters are set the \textit{chief} node starts to change the ``temperature'' for obtaining optimal values in the choice model (lines 10--25 in Algorithm~\ref{alg:chief}). First, it creates a new parameter $\vec{\beta}_{new}$ which is computed as the sum of $\vec{\beta}$ plus a small random error (line 13). Then, it sends $\vec{\beta}_{new}$ to all \emph{workers} so each one can compute and send back the \textit{chief} their own $l_w$ (lines 14--16). In the following step, the \emph{chief} collects and summarizes all $l_w$'s and ``anneals'' the new cost $l_{new}$ with the previous cost $l$ (lines 17--18). If the new cost passes the annealing conditions $\vec{\beta}$, $l$ are updated (lines 20--21), else the parameters are not updated. After a 1000 rounds of ``annealing'', the \emph{chief} change the ``temperature'' (line 25). When the temperature stabilizes, i.e., $temp \leq temp_{min}$ the optimum cost $l$ and $\vec{\beta}$ parameters are obtained. 

\begin{algorithm}
\begin{spacing}{1.1}
\caption{\emph{chief} (dSA)}
\label{alg:chief}
\begin{algorithmic}[1]
\STATE Let be $W$ worker nodes
\STATE send $\vec{\beta}$ to all $W$ \textit{workers}
\STATE Algorithm~\ref{alg:worker} is executed for each \textit{worker}
\STATE get each $l_w$ from each one of the $W$ \textit{workers}  
\STATE $l = \sum_{w=1}^{W} l_w$
\STATE $temp = 1$
\STATE $temp_{min} = 0.00001$
\STATE $\alpha = 0.9$
\STATE $\epsilon = rand(-0.01, 0.01)$ 
\WHILE{$temp > temp_{min}$}
    \STATE $i = 0$
    \WHILE{$i < 1000$}
        \STATE $\vec{\beta}_{new} = \vec{\beta} + \epsilon$
        \STATE send $\vec{\beta}_{new}$ to all $W$ \textit{workers}
        \STATE Algorithm~\ref{alg:worker} is executed for each \textit{worker}
        \STATE get each $l_w$ from each one of the $W$ \textit{workers} 
        \STATE $l_{new} = \sum_{w=1}^{W} l_w$
        \STATE $ap = e^{(l_{new} - l)/temp}$
        \IF{$ap > rand(0,1)$}
            \STATE $\vec{\beta} = \vec{\beta}_{new}$
            \STATE $l = l_{new}$
        \ENDIF
        \STATE $i = i + 1$
    \ENDWHILE
    \STATE $temp = temp * \alpha$
\ENDWHILE
\end{algorithmic}
\end{spacing}
\end{algorithm}

\begin{algorithm}
\begin{spacing}{1.1}
\caption{\emph{worker}-$w$ (dSA)}
\label{alg:worker}
\begin{algorithmic}[1]
\STATE $N$ personal observations
\STATE get $\vec{\beta}$ from the \textit{chief} node
\STATE $l_w = \sum_{n=1}^{N} \ln((y_{a,n}P(a_n;\vec{\beta}) + y_{tr,n}P(tr_n;\vec{\beta})))$
\STATE send $l_w$ to the \textit{chief} node
\end{algorithmic}
\end{spacing}
\end{algorithm}

\subsection{System setup}
We propose the following setup for testing a \emph{maximum likelihood} approach based choice model over BSMD framework. The setup has one \emph{chief}, four \emph{worker} nodes and four additional nodes to run the blockchain and participate in the consensus mechanism. We assume that the \emph{chief} node has enough computing power to run complex computations, while \emph{worker} nodes are individuals that run evaluation of the model on their mobile devices using their personal observations only. Complete instructions for installing the setup can be found at \url{https://github.com/LITrans/saBSMD}. For the case study we use the following configuration:

\begin{enumerate}
    \item[a.] 5 \emph{t3.medium}\footnote{Amazon cloud EC2 virtual machines with 2cores at 3.1GHz and 4GB of RAM} cloud servers. One server simulates the \emph{chief} node and the remaining four are nodes running the blockchain
    \item[b.] 4 Raspberry Pi (RPi) model 3B\footnote{4cores at 1.2GHz and 1GB RAM} using WiFi and running one \emph{worker} node each. The RPIs represent users participating in the distributed choice model using their smartphones.
\end{enumerate}

The \emph{chief} node runs a script similar to Algorithm~\ref{alg:chief} while each RPI runs a script similar to Algorithm~\ref{alg:chief}. For the sake of the study case we randomly divide the $246$ observations of~\cite{Sobhani2017} as follows; three \emph{worker} nodes have $61$ observations and the last one has $64$. 

\subsection{Simulation results}
The dSA algorithm over the BSMD end in about 2hrs and the results are shown in Table~\ref{tab:results}. For verification purposes we solve the \emph{likelihood} based choice model using a non-distributed version of simulated annealing, obtaining the same results as in Table~\ref{tab:results}.

\begin{table}[htbp]
\centering
\caption{Alternative Specific Variables and Constants}
\label{tab:results}
\begin{tabular}{lll}
\hline
\textbf{Parameters} & \textbf{Value} & \textbf{std. err.} \\ \hline
$\beta_{tr}$   & 0 (ref.)       & 0 (ref.)      \\
$\beta_{a}$    & 0.3444         & 0.0012        \\
$\beta_{c}$    & -0.0062        & 0.000016      \\
$\beta_{t}$    & -0.0008        & 0.000003      \\ \hline
\multicolumn{3}{l}{\textit{Model statistics}}             \\
Null \emph{likehood}    & \multicolumn{2}{l}{-257.5839}       \\
Final \emph{likehood}   & \multicolumn{2}{l}{-166.3163}       \\
rho square   & \multicolumn{2}{l}{0.3543}\\ 
\hline
\end{tabular}
\end{table}

Table~\ref{tab:latency} shows the average latency of sending and receiving messages on BSMD for the dSA based estimation process. The latency of \emph{worker} nodes remains under $0.05$ secs, thus the response times are promising for implementing more complex distributed choice models over BSMD. In this particular case, the computational resources needed for running the model on \emph{worker} nodes are minimal. Hence, in theory mobile phone usage is not affected when running the model. However, considering that a node will send and receive approximately $200,000$ messages in the distributed annealing process and that the size of each message is around $400$ Bytes, a total of $80MB$ of information \emph{worker} nodes will send. It is thus recommended that devices use an unmetered network when participating in the distributed annealing process. Fine tuning of the algorithms (e.g. batch evaluation of a set of possible parameters) are also required to minimize the number of messages communicated and iterations needed for conversion.

The average latency of sending a massage from \emph{chief} node is $0.149$ secs, a bit higher than the \emph{workers}. The increment is due to \emph{chief} sending the information to the four \emph{workers} in batches. This could be a problem in a real-life scenario where thousands of \emph{workers} nodes participate, so further exploration is needed regarding this issue. One solution that can be explored is to use multiple \emph{chief} nodes coordinating the distribution estimation process.

\begin{table}[!h]
\centering
\caption{Average latency of sending and getting a message from the BSMD}
\label{tab:latency} 
\begin{tabular}{@{}lcc@{}}
\toprule
\textbf{Nodes}   & \textbf{\begin{tabular}[c]{@{}c@{}}Avg. latency of \\ sending message\end{tabular}} & \textbf{\begin{tabular}[c]{@{}c@{}}Avg. latency of \\ getting message\end{tabular}} \\ \midrule
\textbf{chief}  & 0.149 ($\sigma$ = 0.031)                                                               & 0.043 ($\sigma$ = 0.039)                                                              \\
\textbf{worker 1} & 0.036 ($\sigma$ = 0.020)                                                               & 0.041 ($\sigma$ = 0.028)                                                              \\
\textbf{worker 2} & 0.037 ($\sigma$ = 0.029)                                                               & 0.042 ($\sigma$ = 0.025)                                                              \\
\textbf{worker 3} & 0.036 ($\sigma$ = 0.033)                                                               & 0.043 ($\sigma$ = 0.021)                                                              \\
\textbf{worker 4} & 0.037 ($\sigma$ = 0.026)                                                               & 0.041 ($\sigma$ = 0.022)                                                              \\ \bottomrule
\end{tabular}
\end{table}
\section{Discussion}
In blockchain developments for cryptocurrencies, incentives are used to convince users into participating in the consensus mechanisms, host the network, and trade assets. Hence, blockchain is a perfect platform for the users to control and sell their information, and at the same time companies will benefit from the blockchain as they may be able to exploit a massive market for buying information. In particular the BSMD platform is \textit{public} so everyone can query the ledger and find with whom they are sharing information. Ability to tracking personal data usage brings transparency and a level of confidence that user's data is used for the purpose they agree upon. In the case where a node misuse or leak information, the \textit{public} characteristics of the BSMD will help authorities to easily find all affected parties by the untrusted node and take legal actions.

Identity keys in BSMD create a system of trust, when users participate in the distributed estimation of mobility models. With the identity keys, \emph{chief} nodes can verify the age, gender or other valuable information of \emph{workers} nodes. At the same time \emph{workers} nodes can verify if the \emph{chief} is a reputable or a real organization. Verification minimizes the risk of sharing data with untrusted nodes that could lead to develop poor mobility models or prevent misbehaving nodes scam individuals to participate in fake distributed models. As we mention in Section~\ref{sec:distributed_bsmd} the \emph{smart contracts} are scripts that self-enforce a fair trade in all transactions, these scripts are embedded within the blockchain so it is virtually impossible to alter its contents without other nodes noticing the change. Identity keys and \emph{smart contracts} are tools for transparency that work on both sides and are meant to secure all transactions and protect the privacy of individuals.

By not sharing personal observations our framework protects the identity of participants, while preserving the advantages large-scale data can bring into mobility models. Even if someone intercepts the evaluated objective function value and the model for each node, to the best of our knowledge, the best the attacker could do is to only approximate the personal observations. Also, all transactions are \emph{peer-to-peer} and protected with asymmetric encryption so if an attacker would want to steal evaluated objective functions from various nodes they would have to tamper all connections and decrypt all keys from all users, which may not be worth the effort.

\section{Conclusions and future work}
Protecting individual's privacy is an important issue that has to be tackled if researchers, governments and companies want to fully exploit the potential of large-scale personal data for developing mobility behaviour models. Distributing models and algorithms over BSMD framework is one solution for preserving the identity of individuals in such models. In this paper we present a distributed tool over BSMD for choice modelling where personal raw information is never shared, instead users shares evaluated objective functions given some parameters and model structure.

The proposed framework can work on mobile phones, but more advanced mobility models may consume significant computation power or bandwidth. If the market share of smart phones and/or unmetered networks is low, there exists a risk of creating biased models with the wealthiest sub-populations only. Even though in some developed cities the distributed choice modelling over the BSMD is feasible in terms of population heterogeneity for some other cities this technology is not ready to be implemented yet.

\emph{Chief} nodes knows the model and have access to the decrypted \emph{worker} evaluations, hence if the \emph{chiefs} are \emph{honest but curious} (in the best case) they may be able to infer personal observations of \emph{workers} by using the evaluations. In future, we will consider the use differential privacy techniques for adding noise to the data the \emph{workers} sends and in this manner preventing the \emph{chief} node to infer the data of the \emph{workers}.

In future developments we will implement the distributions of model over the BSMD using more powerful data-driven prediction tools like Federated Learning~\cite{GoogleAI2017} and also adapt advanced Machine Learning algorithms for behavioural models like the one presented in~\cite{wong2018modelling}. Machine Learning is gaining momentum and in combination with blockchain there is a strong potential to develop powerful distributed intelligence, where users can privately and securely train models to create comprehensive prediction tools.  

\newpage
\bibliographystyle{unsrt}  
\bibliography{ITSC_abbre}  

\begin{thebibliography}{10}

\bibitem{farooq2015ubiquitous}
Bilal Farooq, Alexandra Beaulieu, Marwan Ragab, and Viet {Dang Ba}.
\newblock {Ubiquitous monitoring of pedestrian dynamics: Exploring wireless ad
  hoc network of multi-sensor technologies}.
\newblock In {\em 2015 IEEE SENSORS}, pages 1--4. IEEE, nov 2015.

\bibitem{Cadwalladr2018}
Carole Cadwalladr and Emma Graham-Harrison.
\newblock Revealed: 50 million facebook profiles harvested for cambridge
  analytica in major data breach.
\newblock {\em Guardian News and Media Limited}, 2018.

\bibitem{Raghupathi2014}
Wullianallur Raghupathi and Viju Raghupathi.
\newblock {Big data analytics in healthcare: promise and potential}.
\newblock {\em Health Information Science and Systems}, 2(1):3, dec 2014.

\bibitem{Badu-Marfo2019a}
Godwin Badu-Marfo, Bilal Farooq, and Zachary Patterson.
\newblock A perspective on the challenges and opportunities for privacy-aware
  big transportation data.
\newblock {\em Journal of Big Data Analytics in Transportation}, 1(1):1--23,
  Jun 2019.

\bibitem{Amalina2019}
F.~{Amalina}, I.~A.~T. {Hashem}, Z.~H. {Azizul}, A.~T. {Fong}, A.~{Firdaus},
  M.~{Imran}, and N.~B. {Anuar}.
\newblock Blending big data analytics: Review on challenges and a recent study.
\newblock {\em IEEE Access}, pages 1--1, 2019.

\bibitem{Nakamoto2008}
Satoshi Nakamoto.
\newblock Bitcoin: A peer-to-peer electronic cash system.
\newblock Technical report, Bitcoin, 2008.

\bibitem{Bach2018}
L.~M. {Bach}, B.~{Mihaljevic}, and M.~{Zagar}.
\newblock Comparative analysis of blockchain consensus algorithms.
\newblock In {\em 2018 41st International Convention on Information and
  Communication Technology, Electronics and Microelectronics (MIPRO)}, pages
  1545--1550, May 2018.

\bibitem{Lopez2018}
David Lopez and Bilal Farooq.
\newblock {A blockchain framework for smart mobility}.
\newblock In {\em 2018 IEEE International Smart Cities Conference (ISC2)},
  pages 1--7, Kansas City, Missouri, sep 2018. IEEE.

\bibitem{lopez2019multi}
David Lopez and Bilal Farooq.
\newblock A multi-layered blockchain framework for smart mobility data-markets.
\newblock {\em arXiv preprint arXiv:1906.06435}, 2019.

\bibitem{Barff1982}
Richard Barff, David Mackay, and Richard~W. Olshavsky.
\newblock {A Selective Review of Travel-Mode Choice Models}.
\newblock {\em Journal of Consumer Research}, 8(4):370--380, 03 1982.

\bibitem{Hagenauer2017}
Julian Hagenauer and Marco Helbich.
\newblock {A comparative study of machine learning classifiers for modeling
  travel mode choice}.
\newblock {\em Expert Systems with Applications}, 78:273--282, jul 2017.

\bibitem{Ramezani2018}
Samira Ramezani, Barbara Pizzo, and Elizabeth Deakin.
\newblock Determinants of sustainable mode choice in different socio-cultural
  contexts: A comparison of rome and san francisco.
\newblock {\em International Journal of Sustainable Transportation},
  12(9):648--664, 2018.

\bibitem{Pike2018}
Susan Pike and Mark Lubell.
\newblock The conditional effects of social influence in transportation mode
  choice.
\newblock {\em Research in Transportation Economics}, 68:2--10, August 2018.

\bibitem{Halat2015}
Hooram Halat, Meead Saberi, Charlotte~Anne Frei, Andreas~Rolf Frei, and Hani~S.
  Mahmassani.
\newblock Impact of crime statistics on travel mode choice: Case study of the
  city of chicago, illinois.
\newblock {\em Transportation Research Record}, 2537(1):81--87, 2015.

\bibitem{Muzammal2019}
Muhammad Muzammal, Qiang Qu, and Bulat Nasrulin.
\newblock Renovating blockchain with distributed databases: An open source
  system.
\newblock {\em Future Generation Computer Systems}, 90:105 -- 117, 2019.

\bibitem{Li2019}
M.~{Li}, J.~{Weng}, A.~{Yang}, W.~{Lu}, Y.~{Zhang}, L.~{Hou}, J.~{Liu},
  Y.~{Xiang}, and R.~H. {Deng}.
\newblock Crowdbc: A blockchain-based decentralized framework for
  crowdsourcing.
\newblock {\em IEEE Transactions on Parallel and Distributed Systems},
  30(6):1251--1266, June 2019.

\bibitem{Stanciu2017}
A.~{Stanciu}.
\newblock Blockchain based distributed control system for edge computing.
\newblock In {\em 2017 21st International Conference on Control Systems and
  Computer Science (CSCS)}, pages 667--671, May 2017.

\bibitem{Pop2018}
Claudia Pop, Tudor Cioara, Marcel Antal, Ionut Anghel, Ioan Salomie, and
  Massimo Bertoncini.
\newblock Blockchain based decentralized management of demand response programs
  in smart energy grids.
\newblock {\em Sensors}, 18(1), 2018.

\bibitem{Yang2019}
Qiang Yang, Yang Liu, Tianjian Chen, and Yongxin Tong.
\newblock Federated machine learning: Concept and applications.
\newblock {\em ACM Trans. Intell. Syst. Technol.}, 10(2):12:1--12:19, January
  2019.

\bibitem{Olnes2017}
Svein {\O}lnes, Jolien Ubacht, and Marijn Janssen.
\newblock {Blockchain in government: Benefits and implications of distributed
  ledger technology for information sharing}.
\newblock {\em Government Information Quarterly}, 34(3):355--364, sep 2017.

\bibitem{Sobhani2017}
Anae Sobhani and Bilal Farooq.
\newblock Innovative intercity transport mode: Application of choice preference
  integrated with attributes nonattendance and value learning.
\newblock In {\em 21st International Federation of Operational Research
  Societies}, Qu{\'e}b{\'e}c City, Qu{\'e}b{\'e}c, 2017.

\bibitem{ross1990course}
Sheldon~M Ross.
\newblock {\em A course in simulation}.
\newblock Prentice Hall PTR, 1990.

\bibitem{GoogleAI2017}
{Google AI}.
\newblock Federated learning: Collaborative machine learning without
  centralized training data, 2017.

\bibitem{wong2018modelling}
Melvin Wong and Bilal Farooq.
\newblock {Modelling Latent Travel Behaviour Characteristics with Generative
  Machine Learning}.
\newblock In {\em IEEE Conference on Intelligent Transportation Systems,
  Proceedings, ITSC}, volume 2018-November, pages 749--754. IEEE, nov 2018.

\end{thebibliography}

\end{document}